\documentclass{sn-jnl}
\usepackage{amscd}
\usepackage{amsmath}
\usepackage{amssymb}
\usepackage{nicefrac}
\usepackage{hyperref}
\usepackage{graphicx}
\usepackage{mathptmx}
\usepackage{bm}
\DeclareGraphicsExtensions{.eps, .pdf, .jpg, .tif}

\newcommand{\kb}{k_{\text{B}}}
\newcommand{\He}{$^3$He}
\newcommand{\Heb}{$^3$He-B}
\newcommand{\Hea}{$^3$He-A}
\newcommand{\Hefour}{$^4$He}
\newcommand{\Heaero}{$^3$He-aerogel}
\newcommand{\vDel}{\vec{\Delta}}
\def\vz{{\bf z}}
\def\vp{{\bf p}}
\def\hp{\hat\vp}
\def\ns{\negthickspace}
\newcommand{\eps}{\epsilon_n}
\newcommand{\epst}{\widetilde{\epsilon}_n}
\newcommand{\onehalf}{\frac{\mbox{\small 1}}{\mbox{\small 2}}}
\usepackage{natbib}
\setcitestyle{numbers}
\begin{document}
\title{The Heat Capacity of \Heb\ in Silica Aerogel}
\author{J. A. Sauls}
\affil{Hearne Institute of Theoretical Physics, Department of Physics and Astronomy, Louisiana State University, Baton Rouge, LA 70803}

\date{\today}

\abstract{
The thermodynamic potential for superfluid \Heb\ embedded in a homogeneously distributed random potential is calculated from a quasiclassical reduction of the Luttinger-Ward functional to leading order in $\small=\kb T_c/E_f$. The resulting functional provides an extension of the Ginzburg-Landau free energy functional to all temperatures $0<T\le T_c$. Theoretical predictions based on this functional for the heat capacity of superfluid \Heb\ embedded in homogeneous, isotropic silica aerogel are in good agreement with experimental reports for superfluid \Heb\ infused into 98.2\% porous silica aerogel over the pressure range $p=11-29\,\mbox{bar}$. The analysis supports a conclusion that superfluid \Heb\ infused into high-porosity silica aerogels is a gapless superfluid at all pressures.
}



\maketitle


\vspace*{-5mm}
\section{Introduction}

Heat capacity measurements provide key signatures of the transition from the normal to the superfluid phase of liquid \He. In pure \He\ the magnitude of the jump in specific heat provides a direct measure of strong-coupling corrections to weak-coupling BCS theory for the superfluid phases~\cite{rai76}, as well as direct evidence of spin-triplet pairing~\cite{hal76}.
For \He\ confined in sub-micron cavities, or embedded in low-density random solids, scattering of quasiparticles inhibits pair-formation, suppresses both $T_c$ and the discontinuity in the heat capacity, $\Delta C/\gamma_{\text{S}}T_c$, both of which determine the magnitude of the superfluid order paramter in the region near the transition~\cite{thu98,vor03}.
The low temperature heat capacity also reflects the spectrum of quasiparticle states. For pure bulk \Heb\ the quasiparticle spectrum is gapped everywhere on the Fermi surface leading to exponential suppression of the quasiparticle contribution to the specific heat, $C_v \propto e^{-\Delta/\kb T}$~\cite{bal63}.
But for pure bulk \Hea, e.g. stabilized by a magnetic field, the excitation gap has nodes at two symmetry protected points on the Fermi surface, i.e. $\hat\vp=\pm\hat\vz$, $\vert\Delta(\vp)\vert=\Delta\,\left(1-(\hat\vz\cdot\hat{\vp})^2\right)^{1/2}$. Thus, the spectrum is gapless at these two points on the Fermi surface, and the corresponding quasiparticle spectrum at excitation energies well below the maximum gap in the A-phase, $\epsilon\ll \Delta_{\text{0}}$, scales as $N(\epsilon)\propto N_f\,(\epsilon/\Delta_{\text{0}})^2$. This spectrum implies a low-temperature specific heat that is a power law with $C_v \propto (\kb T/\Delta_0)^3$~\cite{vollhardt90}.

For \He\ confined in cavities, or for \He\ films, boundary scattering of quasiparticles is pair-breaking and leads to the formation of sub-gap quasiparticle states that modify the the thermodynamic properties of bulk superfluid \He~\cite{vor03,tsu10}. Evidence from specific heat measurements of subgap quasiparticles (surface Andreev bound states) from surface pair breaking was reported for superfluid \He\ confined in silver powder~\cite{cho06}, and for \He\ confined in a cooper cavity coated with \Hefour\ to provide specular scattering~\cite{bun13}.

Sub-gap quasiparticles are also generated by scattering from the random potential when \He\ is infused into low-density random solids~\cite{sha01}. Indeed specific heat measurements for the B phase of superfluid \He\ confined in high porosity silica aerogel were interpreted to imply the existence of a finite density of quasiparticle states at the Fermi level~\cite{cho04a}.
In this report I present analysis of the specific heat data from Choi et al.~\cite{cho04a} based on a free energy functional of the superfluid order parameter that incorporates pair-breaking from quasiparticle scattering off silica aerogel strands and clusters, and is valid at over the full temperature range below $T_c$. 

\vspace*{-5mm}
\section{Free Energy Functional}

The order parameter of superfluid \He\ in globally isotropic 98.2\% porous silica aerogel is identified as the Balian-Werthamer (BW) state, but modified by the pair-breaking effect of quasiparticle scattering off the random potential over the entire pressure range of the phase diagram~\cite{spr96,bar00,sau05}. 
Here I report results and analysis of the specific heat of superfluid \Heb\ based on a free energy functional which extends the GL theory to all temperatures for superfluid \He\ embedded in an homogeneous isotropic random medium as a model for quasiparticle scattering off the random potential of silica aerogel. 
This functional is based on a reduction of the Luttinger-Ward functional to leading order in the expansion parameters of Fermi-liquid theory~\cite{ali11}. 

In the weak-coupling limit for spatially uniform equilibrium states the functional reduces to
\begin{eqnarray}
\Delta\bar{\Omega}\bigl[\vDel,\vDel^*\bigr] 
&=& N_f V\,
\int\frac{d\Omega_{\hat\vp}}{4\pi}
\Bigg\{ 
-\frac{1}{v_1}\,\vert\vDel(\hp)\vert^2
+
2\pi T\sum_{\eps}^{\varepsilon_c}
\Big[
\left(\vert\epst\vert-\sqrt{\vert\epst\vert^2+\vert\vDel(\hp)\vert^2}\right)
\label{eq-Omega-line1}
\nonumber\\
&&\qquad
\times\left(1+\frac{\hbar}{2\tau\vert\epst\vert}\right)
+
\frac{\hbar}{2\tau}\ln\left(\frac{\sqrt{\vert\epst\vert^2
+
\vert\vDel(\hp)\vert^2}}{\vert\epst\vert}\right)
\Big]
\Bigg\}
\,,
\label{eq-Omega-line2}
\\
\mbox{where}\;\;\epst &=& \eps +
\int\frac{d\Omega_{\hat\vp}}{4\pi}
\Bigg\{ 
\frac{\hbar}{2\tau}
\frac{\sqrt{\vert\epst\vert^2
+
\vert\vDel(\hp)\vert^2}}{\epst}
\Bigg\}
\,,
\label{eq-Omega-line3}
\end{eqnarray}
and $\eps=(2n+1)\pi\kb T$ are the Fermion Matsubara energies. The random potential is parametrized by the quasiparticle-impurity scattering rate $1/\tau$, or mean-free path, $\ell=v_f\tau$. The sum in Eq.~\eqref{eq-Omega-line1} is cutoff for $\vert\eps\vert \ge \varepsilon_c$, where $\kb T_c \ll \varepsilon_c \ll E_f$. The cutoff defining the bandwidth of attraction, $\varepsilon_c$, as well as the strength of the pairing interaction, $v_1$, in the p-wave channel can be eliminated in favor of the measured transition temperature of pure, bulk superfluid \He\ via the eigenvalue of the linearized gap equation, $\kb T_{c_0}=1.13\,\varepsilon_c\,e^{-1/v_1}$. The functional in Eq.~\eqref{eq-Omega-line2} reduces to the weak-coupling BCS free energy functional for pure \He\ (Eq.~5.16 of Ref.~\cite{ser83}) for $\tau\rightarrow\infty$.

Furthermore, the stationarity condition for the impurity-averaged functional, $\delta\Delta\bar\Omega/\delta\vDel(\hp)^*=0$, generates the impurity-renormalized gap equation,
\begin{equation}
\ln\left(\frac{T}{T_{c_0}}\right) 
\vDel(\hp)
\ns=\ns 
\int\frac{d\Omega_{\hat\vp'}}{4\pi}
\Bigg\{ 
\ns
3(\hp\cdot\hp')\,
2\pi\kb T\sum_{n\ge 0}^{\infty}
	\left(
	\ns
	\frac{1}{\sqrt{\epst^2 + \vert\vDel(\hp')\vert^2}}
	-
	\frac{1}{\eps}
	\ns
	\right)
\vDel(\hp')
\Bigg\}
\,.
\end{equation}

\vspace*{-5mm}
\section{Suppression of Pair Formation}

The renormalized Matsubara energy encodes the effect of de-pairing by the scattering of pair-correlated quasiparticles off the distribution of impurities. This is evident from the suppression of the transition temperature~\cite{thu98},
\begin{eqnarray}
\ln\left(\frac{T_c}{T_{c_0}}\right) 
&=& 
\sum_{n\ge 0}^{\infty}
	\left(
	\frac{1}{n+\onehalf + \onehalf\,\alpha\frac{T_{c_0}}{T_c}}
	-
	\frac{1}{n+\onehalf}
	\right)
\,.
\end{eqnarray}
The dimensionless pair-breaking parameter $\alpha$ is the product of the mean scattering rate, $1/\tau$, and the Cooper pair formation time, $\hbar/2\pi\kb T_{c_0}$, i.e. $\alpha=\hbar/2\pi\tau\,\kb T_{c_0}$, The pair-breaking parameter can also be expressed as the ratio of the ballistic coherence length, $\xi_0=\hbar v_f/2\pi k_B T_{c_0}$, of pure \He\ and the transport mean free path, $\ell=v_f\tau$, i.e. $\alpha=\xi_0(p)/\ell$.
The range of disorder from weak to strong pair breaking can be explored by varying the pressure and/or aerogel density. 
The solution for the universal curve for $T_c/T_{c_0}$ versus $\xi_0/\ell$ is shown in Fig.~6 of Ref.~\cite{hal04}. The disorder critical point where $T_c\rightarrow 0$ is $\alpha_c\simeq 0.281$. Since $\xi=\hbar v_f/2\pi\kb T_c\rightarrow\infty$ as $\alpha\rightarrow\alpha_c$ the homogeneous scattering model (HSM) prediction for the critical point is robust. Indeed the disorder driven quantum critical point has been observed for \He\ in 98.2\% porous aerogels,~\cite{mat97} and provides an accurate result for the quasiparticle-impurity mean free path. However, away from the quantum critical point the HSM does not accurately account for $T_c$ vs. pressure for \Heaero\ due to the correlated nature of the random potential for silica aerogels~\cite{thu98}. The effects of correlations on $T_c$ and other observable properties has been studied in Refs.~\cite{sau03,sau05}.
In order to incorporate aerogel correlations and compare the weak-coupling predictions for the heat capacity, $C/T$, with experimental data reported in Ref.~\cite{cho04a}, I fit the pairbreaking parameter, $\alpha$, at each pressure to the measured transition temperature for \Heaero.
Fermi liquid data, c.f. $N_f=3n/4E_f$, $E_f=\onehalf v_f p_f$, $v_f$, $p_f = \hbar k_f$, $n = k_f^3/3\pi^2$, are obtained from the tabulated data of Ref.~\cite{hal90}. In addition, I used the published value for the volume of \He\ in the aerogel sample of Ref.~\cite{cho04a}, $V_a = 1.028\,\mbox{cm}^3$. 
The resulting normal-state value of $C/T$ agrees nearly perfectly with the reported values in Ref.~\cite{cho04a} except for a slight discrepancy at $p=20.06\,\mbox{bar}$ as shown in 
Fig.~\ref{fig-CoverT-20bar}.

\vspace*{-5mm}
\section{Results for $C_v/T$}

\begin{figure}[h!]
\includegraphics[width=0.90\linewidth,keepaspectratio]{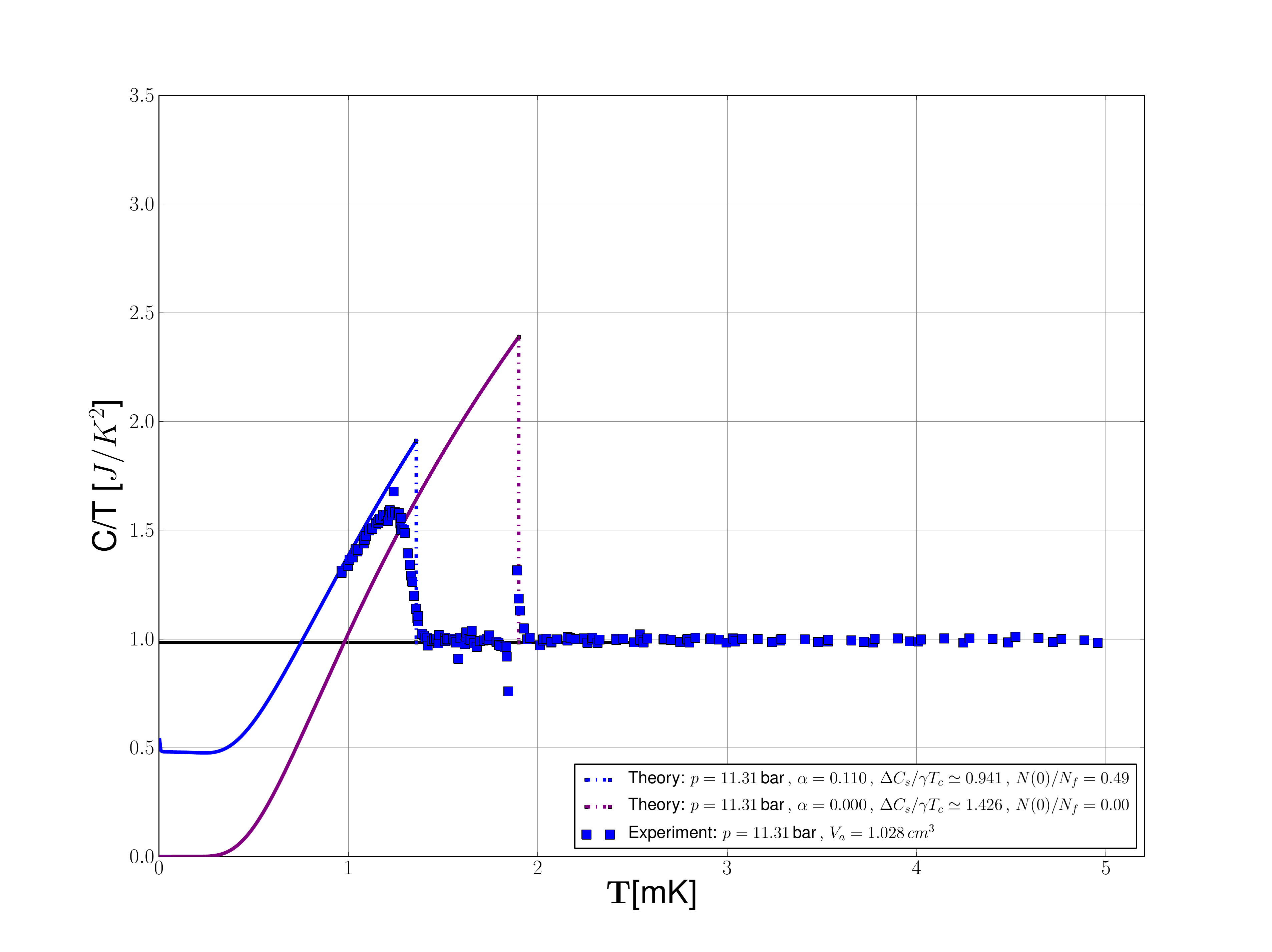}
\caption{Heat Capacity of \Heaero\ at $p = 11.31\,\mbox{bar}$. Theoretical results for the weak-coupling BW phase are shown for pure (maroon) and dirty (blue) \Heb. N.B. subtraction of the bulk heat capacity leaves a trace marking the bulk transition temperature, $T_{c_0}$.}
\label{fig-CoverT-11bar}
\end{figure}

\begin{figure}[H]
\includegraphics[width=0.90\linewidth,keepaspectratio]{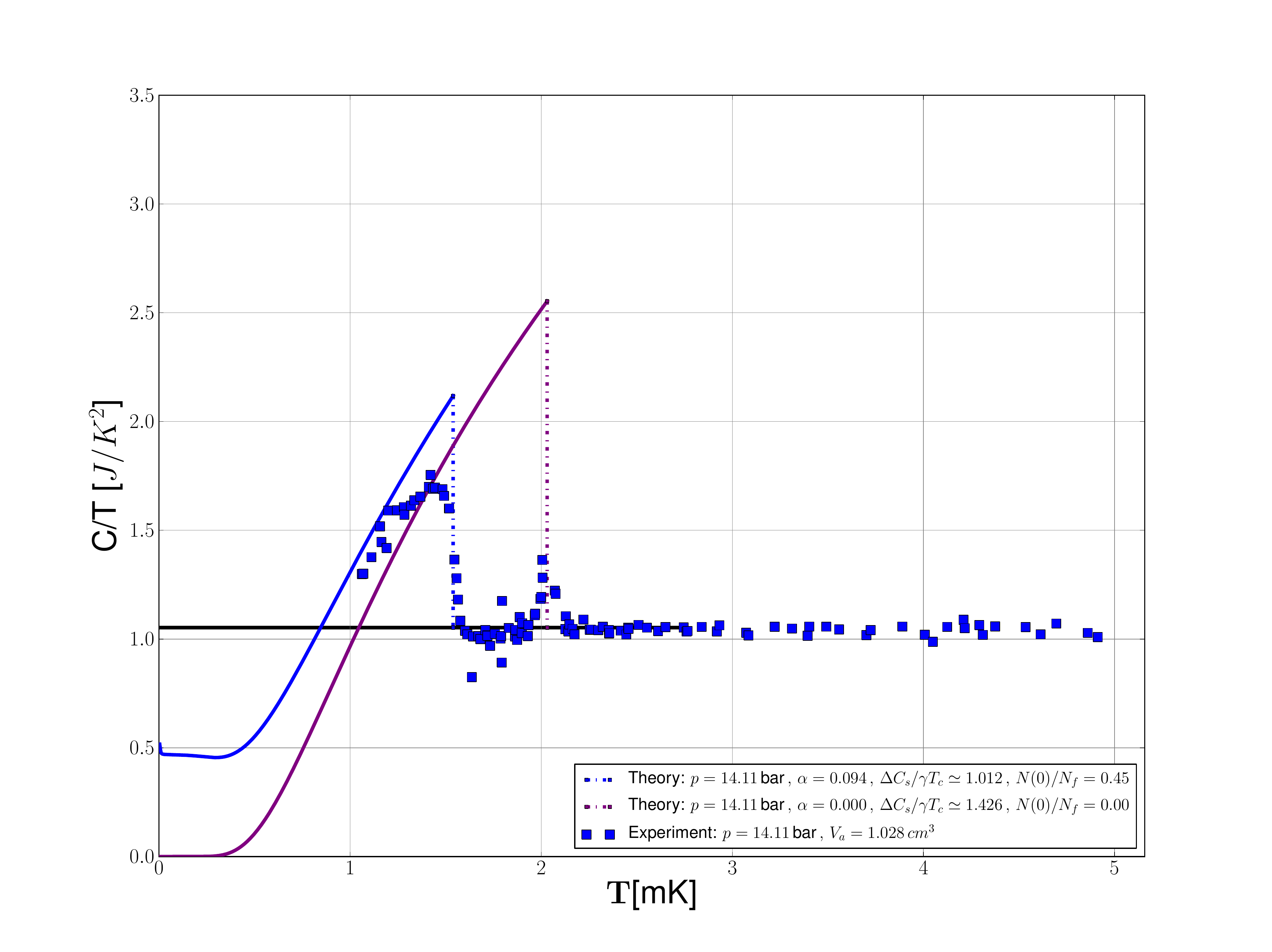}
\caption{Heat Capacity of \Heaero\ at $p = 14.11\,\mbox{bar}$. Theoretical results for the weak-coupling BW phase are shown for pure (maroon) and dirty (blue) \Heb.}
\label{fig-CoverT-14bar}
\end{figure}

\begin{figure}[h!]
\includegraphics[width=0.90\linewidth,keepaspectratio]{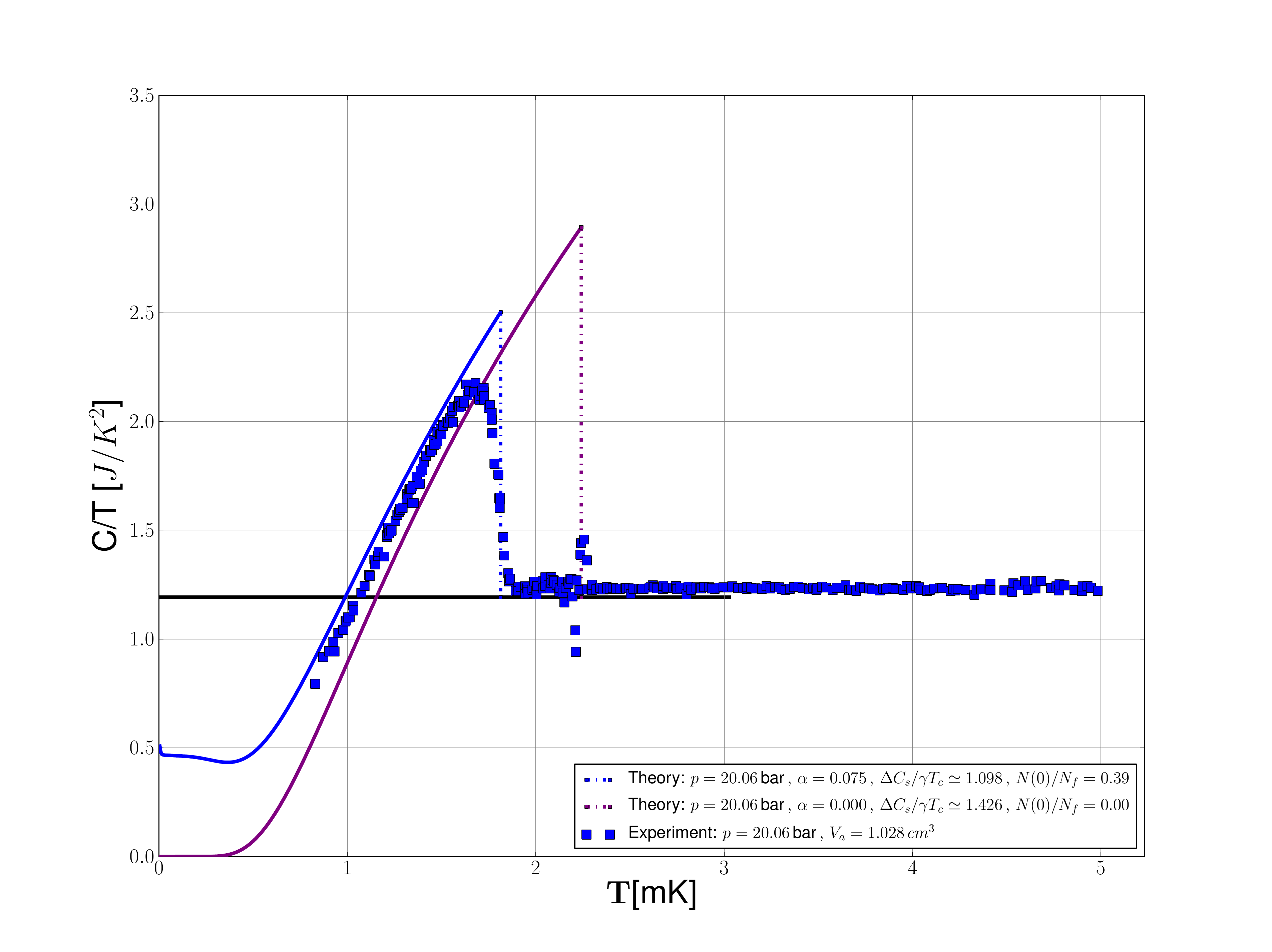}
\caption{Heat Capacity of \Heaero\ at $p = 20.06\,\mbox{bar}$. Theoretical results for the weak-coupling BW phase are shown for pure (maroon) and dirty (blue) \Heb.}
\label{fig-CoverT-20bar}
\end{figure}

\begin{figure}[h!]
\includegraphics[width=0.90\linewidth,keepaspectratio]{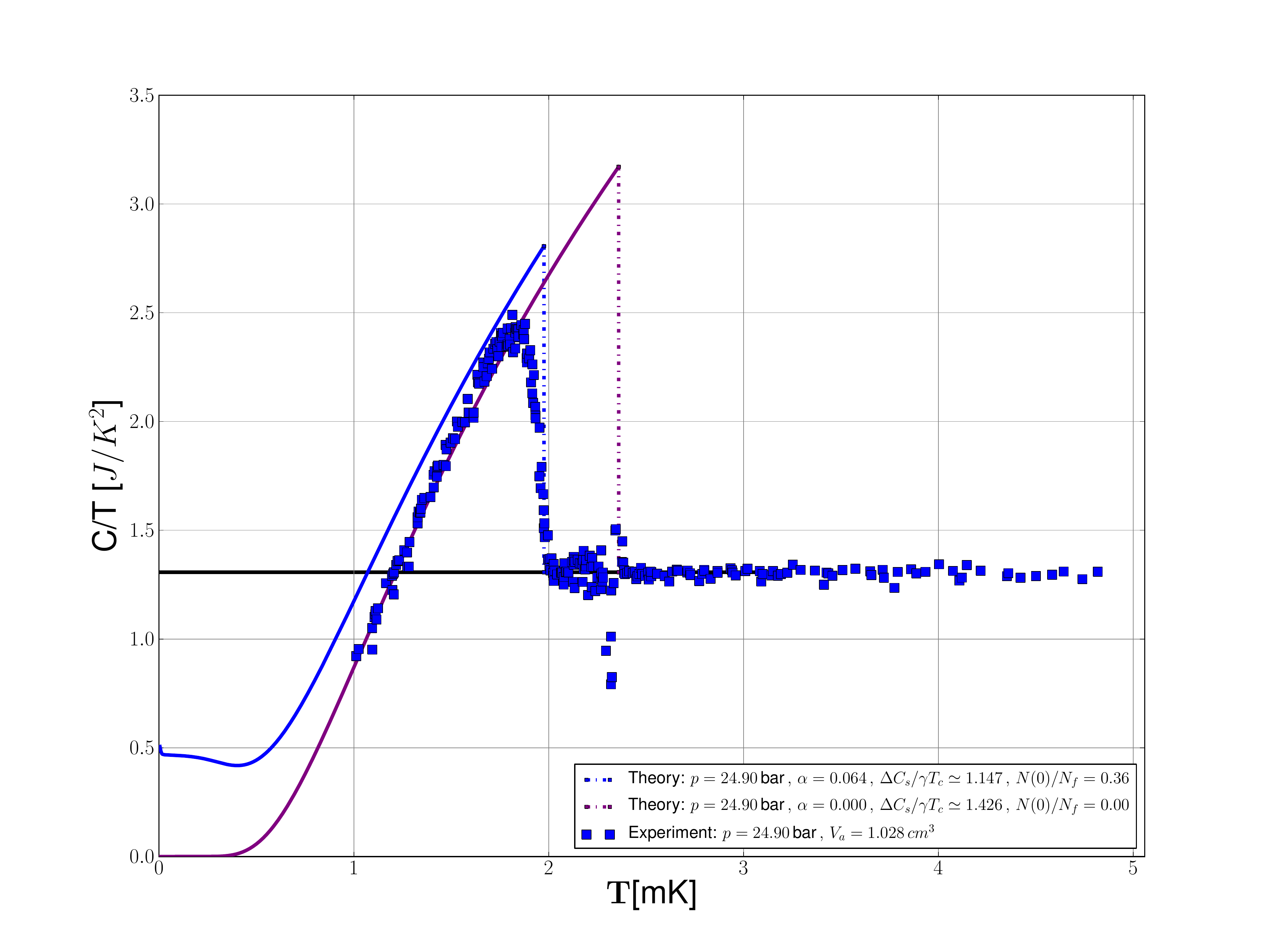}
\caption{Heat Capacity of \Heaero\ at $p = 24.90\,\mbox{bar}$. Theoretical results for the weak-coupling BW phase are shown for pure (maroon) and dirty (blue) \Heb.}
\label{fig-CoverT-24bar}
\end{figure}

\begin{figure}[h!]
\includegraphics[width=0.90\linewidth,keepaspectratio]{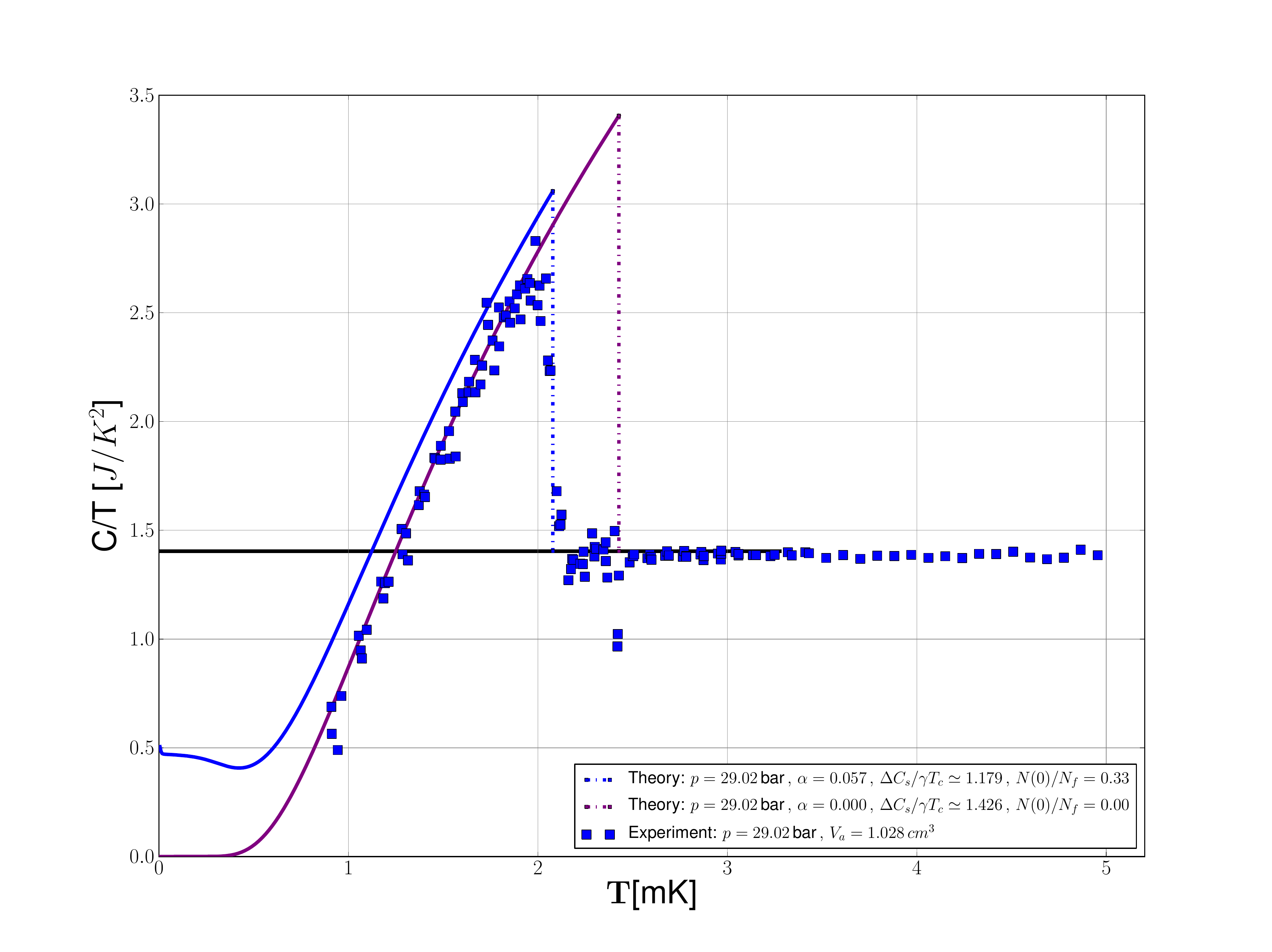}
\caption{Heat Capacity of \Heaero\ at $p = 29.02\,\mbox{bar}$. Theoretical results for the weak-coupling BW phase are shown for pure (maroon) and dirty (blue) \Heb.}
\label{fig-CoverT-29bar}
\end{figure}

One might expect reasonable agreement between theory and experiment for the heat capacity of the superfluid phase at low pressures since both strong-coupling effects and deviations from the HSM are relatively small. Indeed the theoretical calculations of $C/T$ at $p=11.31$, $14.11$ and $20.06$ bar (Figs.~\ref{fig-CoverT-11bar},~\ref{fig-CoverT-14bar} and~\ref{fig-CoverT-20bar}) are all in good agreement with the experimenal data, although there is broadening of the transition visible at all pressures which I expect is associated with large-scale inhomogeneity of the aerogel sample. 
At higher pressures, $p=24.90$ bar and $p=29.02$ bar (Figs.~\ref{fig-CoverT-24bar} and~\ref{fig-CoverT-29bar}), the slope of $C/T$ is still reasonably well reproduced, but there is an increase in the theoretical value of $C/T$ compared to the experimental data.
This shift almost certainly reflects limitations of the HSM based on point impurities. The correlated nature of the aerogel structure, as well as the finite size of the silica strands and clusters, become more important at higher pressures where the Cooper pair correlation length is smallest and approaches the aerogel correlation length.
Note that strong-coupling enchancement of the experimental heat capacity jump is expected to be largest at high pressures, but is suppressed significantly relative to that of pure bulk \He\ at the same pressure.

\begin{figure}[h!]
\includegraphics[width=0.90\linewidth,keepaspectratio]{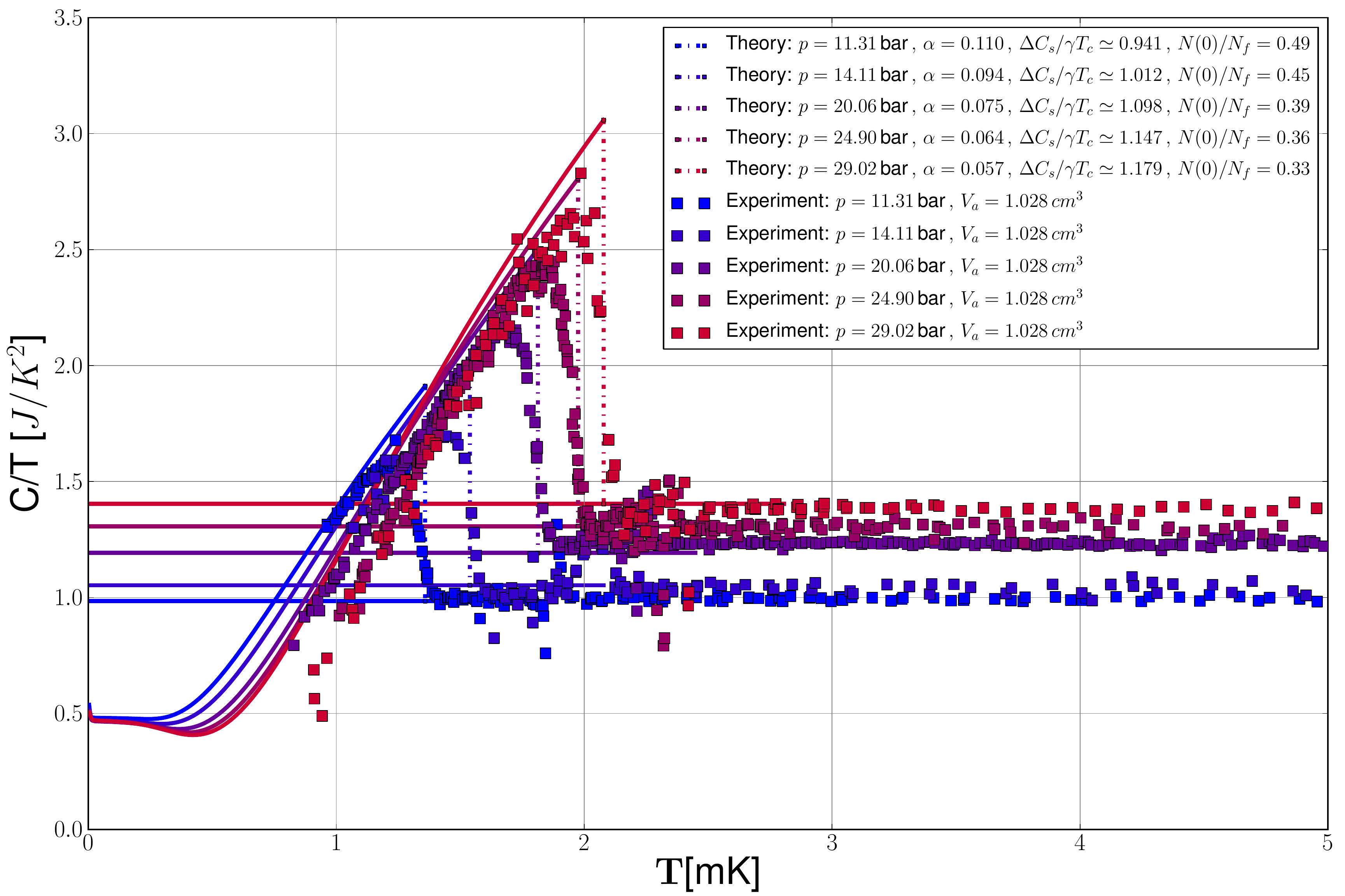}
\caption{Comparison of the Heat Capacity of \Heaero\ for pressures:
$p=11.31\,\mbox{bar}$, $14.11\,\mbox{bar}$, $20.06\,\mbox{bar}$, $24.90\,\mbox{bar}$, $29.02\,\mbox{bar}$. Note the pressure independent value for $\lim_{T\rightarrow 0} C/T\propto N(0)$, in contrast to the pressure dependence of $C/T$ for $T>T_c$.
}
\label{fig-CoverT-all}
\end{figure}

Finally, I note that the density of states at the Fermi level for each pressure, extracted from $\lim_{T\rightarrow 0}C/T\sim N(0)$, is \emph{predicted} to be {\sl independent of pressure} (see Fig.~\ref{fig-CoverT-all}). This is in contrast to the density of states for normal \He\ which varies substantially with pressure as shown in Fig.~\ref{fig-CoverT-all}.   
For the superfluid phase $N(0)$ reflects the density of sub-gap states confined on the aerogel strands and clusters. Thus, it seems reasonable to assume that the values of $N(0)$ are determined by the surface area of aerogel strands that host surface Andreev bound states, and thus is insensitive to pressure and \He\ density. I also note that the non-monotonic behavior of $C/T$ that is predicted at higher pressures is due to the finite bandwidth of the impurity-induced DOS centered at the Fermi energy~\cite{ali11}.

\vspace*{-5mm}
\section{Conclusions}

Superfluid \Heb\ embedded in high-porosity aerogel is predicted to exhibit gapless behavior at low temperatures over the entire pressure range. 
Analysis of the specific heat data for pressures $11\,\mbox{bar} \le p \le 29\,\mbox{bar}$ points directly to a band of sub-gap quasiparticle bound states with a finite density of states at the Fermi level.  
Measurements of $C/T$ on these samples down to $300\,\mu\mbox{K}$ could provide definitive and direct measurements of a finite density of surface Andreev bound states confined on the aerogel structure.  

\vspace*{-5mm}
\subsubsection*{Acknowledgments} 

This research was supported by National Science Foundation Grant DMR-1508730. I thank Bill Halperin and Hyoungsoon Choi for discussions on their experiments and for providing me with their published heat capacity data.


\begin{thebibliography}{10}
\providecommand{\url}[1]{{#1}}
\providecommand{\urlprefix}{URL }
\expandafter\ifx\csname urlstyle\endcsname\relax
  \providecommand{\doi}[1]{DOI \discretionary{}{}{}#1}\else
  \providecommand{\doi}{DOI \discretionary{}{}{}\begingroup
  \urlstyle{rm}\Url}\fi

\bibitem{rai76}
D.~Rainer, J.W. Serene, Phys. Rev. B \textbf{13}, 4745 (1976).
\newblock \doi{10.1103/PhysRevA.10.2386}

\bibitem{hal76}
W.P. Halperin, C.N. Archie, F.B. Rasmussen, T.A. Alvesalo, R.C. Richardson,
  Phys. Rev. B \textbf{13}, 2124 (1976).
\newblock \doi{10.1103/PhysRevB.13.2124}

\bibitem{thu98}
E.V. Thuneberg, S.K. Yip, M.~Fogelstr\"om, J.A. Sauls, Phys. Rev. Lett.
  \textbf{80}, 2861 (1998).
\newblock \doi{10.1103/PhysRevLett.80.2861}

\bibitem{vor03}
A.~Vorontsov, J.A. Sauls, Phys. Rev. B \textbf{68}, 064508 (2003).
\newblock \doi{10.1103/PhysRevB.68.064508}

\bibitem{bal63}
R.~Balian, N.R. Werthamer, Phys. Rev. \textbf{131}, 1553 (1963).
\newblock \doi{10.1103/PhysRev.131.1553}

\bibitem{vollhardt90}
D.~Vollhardt, P.~W\"olfle, \emph{{The Superfluid Phases of $^3$He}} (Taylor \&
  Francis, New York, 1990)

\bibitem{tsu10}
Y.~Tsutsumi, T.~Mizushima, M.~Ichioka, K.~Machida, J. Phys. Soc. Jpn.
  \textbf{79}(11), 113601 (2010).
\newblock \doi{10.1143/JPSJ.79.113601}

\bibitem{cho06}
H.~Choi, J.P. Davis, J.~Pollanen, W.P. Halperin, Phys. Rev. Lett.
  \textbf{96}(12), 125301 (2006).
\newblock \doi{10.1103/PhysRevLett.96.125301}

\bibitem{bun13}
Y.M. Bunkov, J. Low Temp. Phys. \textbf{174}, 1 (2013).
\newblock \doi{10.1007/s10909-013-0933-3}

\bibitem{sha01}
P.~Sharma, J.A. Sauls, J. Low Temp. Phys. \textbf{125}, 115 (2001).
\newblock \doi{10.1023/A:1012913717533}

\bibitem{cho04a}
H.~Choi, K.~Yawata, T.~Haard, J.~Davis, G.~Gervais, N.~Mulders, P.~Sharma,
  J.~Sauls, W.~Halperin, Phys. Rev. Lett. \textbf{93}(14), 145301 (2004).
\newblock \doi{10.1103/PhysRevLett.93.145301}

\bibitem{spr96}
D.~Sprague, T.~Haard, J.~Kycia, M.~Rand, Y.~Lee, P.~Hamot, W.~Halperin, Phys.
  Rev. Lett. \textbf{77}, 4568 (1996)

\bibitem{bar00}
B.I. Barker, Y.~Lee, L.~Polukhina, D.~D.Osheroff, L.W. Hrubesh, J.F. Poco,
  Phys. Rev. Lett. \textbf{85}, 2148 (2000)

\bibitem{sau05}
J.A. Sauls, Y.M. Bunkov, E.~Collin, H.~Godfrin, P.~Sharma, Phys. Rev. B
  \textbf{72}(2), 024507 (2005).
\newblock \doi{10.1103/PhysRevB.72.024507}

\bibitem{ali11}
S.~Ali, L.~Zhang, J.A. Sauls, J. Low Temp. Phys. \textbf{162}(3-4), 233 (2011).
\newblock \doi{10.1007/s10909-010-0310-4}

\bibitem{ser83}
J.W. Serene, D.~Rainer, Phys. Rep. \textbf{101}, 221 (1983).
\newblock \doi{10.1016/0370-1573(83)90051-0}

\bibitem{hal04}
W.P. Halperin, J.A. Sauls, arXiv \textbf{cond-mat.supr-con}, 0408593 (2004).
\newblock \urlprefix\url{http://arxiv.org/abs/cond-mat/0408593v1}.
\newblock 10 pages with 12 figures

\bibitem{mat97}
K.~Matsumoto, J.V. Porto, L.~Pollack, E.N. Smith, T.L. Ho, J.M. Parpia, Phys.
  Rev. Lett. \textbf{79}, 253 (1997).
\newblock \doi{10.1103/PhysRevLett.79.253}

\bibitem{sau03}
J.A. Sauls, P.~Sharma, Phys. Rev. B \textbf{68}, 224502 (2003).
\newblock \doi{10.1103/PhysRevB.68.224502}

\bibitem{hal90}
W.P. Halperin, E.~Varoquaux, \emph{Order {P}arameter {C}ollective {M}odes in
  {S}uperfluid {{$^3He$}}} (Elsevier Science Publishers, Amsterdam, 1990), p.
  353
\end{thebibliography}
%

\vspace*{-2mm}

\vfill
\eject

\subsubsection*{Figures} 
\vspace*{-8mm}

\vspace*{-10mm}


\end{document}